# ROBOTIC PROCESS AUTOMATION AS A DRIVER FOR SUSTAINABLE INNOVATION AND ENTREPRENEURSHIP

Petr Průcha[0000-0003-2197-7825]

Technical University of Liberec, Liberec, Czechia
`petr.prucha@tul.cz`

**Abstract:**

Technological innovation plays a crucial role in driving economic growth and development. In this study, we investigate the extent to which technological innovation contributes to a more sustainable future and fosters entrepreneurship. To examine this, we focus on robotic process automation (RPA) highly relevant technology. We conducted a comprehensive analysis by examining the usage of RPA and its impact on environmental, social, and governance (ESG) factors. Our research involved gathering data from the 300 largest companies in terms of market capitalization. We assessed whether these companies used RPA and obtained their corresponding ESG ratings. To investigate the relationship between RPA and ESG, we employed a contingency table analysis, which involved categorizing the data based on ESG ratings. We further used Pearson's Chi-square Test of Independence to assess the impact of RPA on ESG. Our findings revealed a statistically significant association between RPA and ESG ratings, indicating their interconnection. The calculated value for Pearson's Chi-square Test of Independence was 6.54, with a corresponding p-value of 0.0381. This indicates that at a significance level of five percent, the RPA and ESG variables depend on each other. These results suggest that RPA, representative of modern technologies, likely influences the achievement of a sustainable future and the promotion of entrepreneurship. In conclusion, our study provides empirical evidence supporting the notion that technological innovations such as RPA have the potential to positively shape sustainability efforts and entrepreneurial endeavours.

## 1 INTRODUCTION

Technological innovation has become a driving force behind economic growth and development, with its potential to shape a more sustainable future and foster entrepreneurship being widely recognized. One such cutting-edge technology that has garnered significant attention is robotic process automation (RPA). RPA offers the promise of automating routine and monotonous tasks, but its impact on the environment, society, and governance (ESG) factors remains a subject of debate and uncertainty in the research community (Syed et al., 2020; Wewerka & Reichert, 2021). While some argue that RPA may pose a threat to human employment by potentially overtaking certain job functions, others contend that it complements human capabilities, freeing individuals from mundane tasks and enabling them to engage in more creative and cognitively demanding work (Willcocks et al., 2015). Additionally, the cost-saving potential of RPA in reducing human labour expenses adds another layer of complexity to the discussion (Eikebrokk & Olsen, 2020).

In this work, we want to focus on whether robotic process automation contributes to environmental, social, or corporate governance. We want to investigate whether there is any connection between companies using or not using RPA and their environmental, social and governance responsibility. The overarching aim is to see if technologies such as RPA have any impact on ESG.

Current work on RPA and ESG is limited to case studies and beneficial use cases of employing RPA, especially in the medical field, where RPA has positive impacts on people's lives by allowing faster data flow across hospitals and laboratories to expedite patient treatment and release from hospitals (Hewitt et al., 2021). Unfortunately, there is no more work on robotic process automation and environmental, social, and corporate governance in the databases of the Web of Science and Scopus. We searched for similar studies on technologies and ESG rating in WoS and Scopus databases. ESG rating helps to promote innovation in companies, not only technological innovation, but also ethical and especially green innovation to reduce the carbon foot print in production (Zhang & Jin, 2022). Saxena et al., (2022) mention that there is a demand for analysis into how trending technologies such as blockchain, artificial intelligence, digital twins, automation, etc. influence the ESG score and how to use modern technologies to perform ESG analyses more frequently. In a study by Macpherson et al., (2021), they mention using technologies such as artificial intelligence or Fin-Tech to calculate ESG rating.

We could not find a direct connection between RPA and ESG in any research article. However, there are a few indirect connections in works related to each aspect of ESG rating. For the environmental aspect, we located studies on how RPA helps save energy by automating energy-saving operations in buildings (Yamamoto et al., 2020). For better management of energy usage, RPA could be implemented as a parameter in scheduling proposals (Seguin & Benkalai, 2020), which would consider peaks of electricity in electrical networks and schedule them for off-peak hours. This would reduce the electricity cost of using RPA. With safe and secure cloud infrastructure, RPA could run in cloud centers leveraging the possibility of better electricity management (Chiaraviglio et al., 2018). For social aspects of robotic process automation, we took a broader perspective and searched for social aspects of human and robot cooperation in the workplace. Humans and robots have been working together for centuries with clear expectations from all stakeholders, and the collaborations were historically beneficial. With a higher complexity of technologies, it is important to carefully explain and set expectations with all stakeholders who are involved in the implementation of human and robot collaboration for beneficial purposes (Ajoudani et al., 2018; Sauppé & Mutlu, 2015).

The governance aspect of robotic process automation is connected to the social aspect of providing clear explanations. RPA as a tool can bring a higher level of transparency into organizations and can help manage all legal activities (Lacity & Willcocks, 2016; Syed et al., 2020; Wewerka & Reichert, 2021). It is up to the people in organizations to leverage the benefits of RPA technology and ensure the moral integrity of organizations (Buchholz & Rosenthal, 2002). Because there is no existing research drawing a direct correlation between RPA and ESG, we will investigate the topic in this article. We examine whether RPA, as a modern form of technology, influences ESG.

Firstly, in this paper, we state the problem and the objective of the research. Next, we introduce the research methods. Following that, we will present the results along with the discussion and implications derived from our research. Finally, we conclude our study.

## 2 METHODS OF RESEARCH

To investigate the impact of using RPA in companies on their ESG performance, we selected the top 300 companies in the world by market capitalization. The 300 companies were selected to date on March 15th, 2023 on the website: companiesmarketcap.com. For these selected companies, we investigated whether any RPA technology is used. We searched any public record on their websites or vendor websites or we examined whether the company searches for jobs related to RPA such as: RPA developer, RPA analyst, or RPA engineer. We also examined whether anyone works in the company as an RPA developer, RPA architect, RPA analyst, or engineer. With this procedure, we discovered that, of the top 300 companies in the world, 256 use RPA (see table 1). If we could not find any records about RPA in publicly available data, we reasoned that these companies were not using RPA technology.

Table 1: Statistics about RPA in the top 300 companies

| The number of companies using RPA | The number of companies not using RPA |
|---|---|
| 256 | 44 |

Source: Author's data

For the analysis of companies' environmental, social and corporate governance responsibility, we used ESG ratings from rating agencies. We used ESG rating to capture the complex problematic of ESG. Mostly, we used data from Yahoo Finance and Morningstar for ESG ratings. Except for Hikvision and United Heavy Machinery, we found the ratings for all companies in the list. We were also unable to find any mention of RPA for these two companies, which is likely due to the countries where the companies are located, political issues in these countries and the field in which they do business. For analytic purposes, we used the last known ESG score from 2019 for Hikvision and for United Heavy Machinery, we used the average of the industry standard in the country. For the collection of all the data, we mostly used publicly available data and data sources that are not behind any paywall. Aggregated data by continent and their statistics appear in Table 2. We from collected data we create dataset that is publicly available[1].

Table 2: Aggregated data by continent

| Continent | ESG Average | The number of companies in the continent | The number of companies using RPA | The number of companies not using RPA |
|---|---|---|---|---|
| Asia | 26.7 | 72 | 51 | 21 |
| Australia | 24.6 | 5 | 5 | 0 |
| Europe | 20.4 | 70 | 59 | 11 |
| North America | 21.9 | 150 | 138 | 12 |
| South America | 33 | 3 | 3 | 0 |

Source: Author's data

The ESG rating can vary from 0 to 100. The selected companies fall into a range from 5 to 60. This is due to the fact that they are the top 300 largest companies and most of them are publicly traded; therefore, they need to at least portray a certain image in ESG. For the analysis of dependency, we used a contingency table, viz. table 3 and Pearson's Chi-square Test of Independence (1).

Table 3: Contingency table

| RPA | ESG rating | | |
|---|---|---|---|
| | <0 – 20) Low | <20-40) Medium | <40-60) High |
| Yes | 109 | 139 | 8 |
| No | 15 | 24 | 5 |

Source: Author's work

---

[1] The dataset is publicly available at GitHub: https://github.com/Scherifow/TOP300-Companies-RPA-ESG or Zenodo DOI: 10.5281/zenodo.10615111

We appropriately divided the companies by ESG score into different groups (low, medium and high). We performed the analysis of independence with Anaconda (Python distribution for scientific computing), with a Jupyter notebook, with Python 3.9.11 and the Scipy library for statistics.

$$X^2 = \sum_{i=1}^{n} \frac{(O_i - E_i)^2}{E_i} \qquad (1)$$

## 3 RESULTS

The Pearson's Chi-square test of independence yielded a value of 6.54, indicating a significant relationship between the variables of RPA and ESG rating (p = 0.0381). Based on a significance level of 5 %, we can conclude that there is dependence between these variables. Thus, the results provide evidence that RPA has an influence on environmental, social, and governance factors, although there is a 5 % margin of uncertainty associated with this relationship.

## 4 DISCUSSION

The results from the Pearson's Chi-square Test of Independence show that the ESG rating and RPA technology are not independent variables. This co-relation can be caused by current trends in society; robotic process automation is one of the trending technologies (*Gartner, 2023*). ESG is also trending and there is more initiative to support ESG; consider, for example, trends like ESG investing, which promotes investment exclusively in companies that are ESG responsible (Egorova et al., 2022; Halbritter & Dorfleitner, 2015), or the EU with its green deal forcing banks to offer loans only to ESG responsible companies (Miroshnichenko & Mostovaya, 2019). All these initiatives are related to generations Y & Z, who now constitute the majority of the labour market. Generally, they do not want to work for companies that are not ESG responsible, creating a challenge for those companies to find new employees (Schroth, 2019). The corelation can be linked to the current reality of these trends and the companies selected. All the companies are large and well known and probably none of them need negative publicity related to ESG. Any negative publicity related to ESG potentially leading to a mass boycott from customers would harm their value because the majority of the companies selected are publicly traded rendering negative publicity highly undesirable. Stemming from public opinion and investor pressure, the companies are forced to innovate and use new technologies such as RPA. Robotic process automation is used worldwide because of the benefits it provides including the automation of routine work, fast development of IT even with legacy systems, ability to work 24/7, ease of automating any process, retrieval of human errors from the process, standardization of the process and saving money (Aguirre & Rodriguez, 2017; Anagnoste, 2017; Eikebrokk & Olsen, 2020; Syed et al., 2020). The correlation can be found between the usefulness of incorporating RPA and the initiatives and pressures compelling companies to improve their ESG.

It is also important to mention that the P-value of 0.0381 is really close to the threshold of 0.05 so, on a 1% level of significance, we cannot say that the variables RPA and ESG rating are dependent. We are aware of this limitation. Also, it is important to mention that the dataset is made up of the top 300 companies by market capitalization. If we selected different criteria like the number of employees or the top 500 companies, the results could be different because of different companies in the dataset. It is important to keep in mind that the dependence does not mean causality, and we certainly cannot say whether the RPA causes better or worse ESG ratings. The investigation of the causality of RPA on ESG rating is a potential for future research along with investigations into the impact other technological innovations have on ESG rating.

# 5 CONCLUSION

The primary objective of this study was to examine the influence of robotic process automation (RPA) on the environmental, social, and governance (ESG) responsibility of the top 300 global companies, as determined by market price. To investigate this relationship, a contingency table was constructed, and the Pearson's Chi-square test of independence was employed. The findings indicate that at a significance level of 5%, a statistically significant association between robotic process automation and environmental, social, and governance exists. Therefore, it is reasonable to posit a connection between RPA and ESG.

**ACKNOWLEDGMENT**

This research was made possible thanks to the Technical University of Liberec and the SGS grant number: SGS-2023-1328.

*Contact information (author/s, consultant, corresponding author):*

Petr Průcha

Technical University of Liberec, Faculty of Economics

Studentská 1402/2, Liberec, 461 17

E-mail: petr.prucha@tul.cz